\documentclass[aps,prd,amsmath,amssymb,showpacs]{revtex4}
\usepackage{epsfig,float}
\usepackage{graphicx}
\usepackage{dcolumn}
\usepackage{morefloats}

\usepackage{slashed}
\usepackage{bbm}

\begin{document}

\title{Systematic study of the singularity mechanism in heavy quarkonium decays}

\author{Qian Wang$^1$\footnote{{\it Email address:} q.wang@fz-juelich.de},
Christoph Hanhart$^1$\footnote{{\it Email address:}
        c.hanhart@fz-juelich.de}, Qiang Zhao$^2$\footnote{{\it Email address:}
        zhaoq@ihep.ac.cn} }

\affiliation{
       $^1$ Institut f\"{u}r Kernphysik,  Institute for Advanced Simulation and J\"{u}lich Center for Hadron Physics,
          Forschungszentrum J\"{u}lich, D--52425 J\"{u}lich, Germany\\
       $^2$ Institute of High Energy Physics and Theoretical Physics Center for Science Facilities,
        Chinese Academy of Sciences, Beijing 100049, China}

\begin{abstract}
We investigate in detail the role of heavy meson loops in the
transition from $J^{PC}=1^{--}$ sources to  candidates for QCD
``exotics", such as $Z_c(3900)$, $Z_b(10610)$ and $Z_b'(10650)$.
We demonstrate that, if a vector state strongly couples to a heavy
meson pair in an $S$--wave and this system decays to another
heavy meson pair (e.g. via pion emission), again in an $S$-wave,
the pertinent diagrams get enhanced significantly, if the intermediate states are
(near) on--shell and have small relative momenta.
In a limited kinematic range this mechanism generates ``singularity regions"
that lead to the creation of a large number of low energy heavy meson
pairs, providing an ideal environment
 for the formation of hadron-hadron bound states or
resonances. For instance, we predict that the signals for $Z_b$ and
$Z_b'$ should be a lot stronger in $\Upsilon(6S)$ decays due to
this mechanism, if these states are indeed
hadron-hadron resonances.
 The findings of this work should be valuable for deepening our
understanding of the nature of the mentioned states.
\end{abstract}

\date{\today}


\maketitle

\section{Introduction}

The recent results from the BESIII~\cite{Ablikim:2013mio} and Belle
Collaboration~\cite{Liu:2013dau} have attracted immediate attention
from the hadron physics community. The observation of an enhancement
$Z_c(3900)$ with charge in the invariant mass spectrum of
$J/\psi\pi^\pm$ in $e^+e^-\to Y(4260)\to J/\psi\pi^+\pi^-$ with high
statistics may be a clear evidence for QCD ``exotics" in the
charmonium energy region. The observation is also confirmed by the
CLEO-c experiment in $e^+e^-\to \psi(4170)\to J/\psi\pi\pi$ in the
invariant mass of $J/\psi\pi$~\cite{Xiao:2013iha}. The mass of
$Z_c(3900)$ is close to the $\bar DD^*$ threshold. It therefore is
an interesting analogue to $Z_b(10610)$ and $Z_b(10650)$, located
close to the $\bar BB^*$ and $\bar B^*B^*$ thresholds, respectively,
which were observed by the Belle
Collaboration~\cite{Belle:2011aa,Adachi:2012im} last year. There
have been a lot of theoretical efforts on the interpretation of the
$Z_b$
states~\cite{bondar,Voloshin:2013ez,Li:2012uc,Li:2012as,Ohkoda:2012rj,Cleven:2011,Cleven:2013sq,bugg,mehen,Chen:2012yr,Chen:2011pv}.
Almost immediately after BESIII published their data, different
interpretations~\cite{Wang:2013cya,Guo:2013sya,Chen:2013wca,Faccini:2013lda,
Karliner:2013dqa,Voloshin:2013dpa,bonn,Mahajan:2013qja,Cui:2013yva,Li:2013xia,Zhang:2013aoa,Chen:2013coa,Dias:2013xfa}
were proposed for understanding the nature of the $Z_c(3900)$.

In Ref.~\cite{Wang:2013cya} it was argued that, if  there is a
significant amount of $\bar DD_1+c.c.$ in the wave function of
$Y(4260)$, namely, if the $Y(4260)$ is predominantly of molecular
nature, then a large number of low energy $L=0$ $\bar DD^*$ pairs
would be naturally produced, since both  $\bar DD_1$ and $\bar DD^*$
can be nearly on-shell in a relative $S$-wave simultaneously. This
leads to a significant enhancement of the pertinent loops and
provides an ideal environment for the formation of  $\bar DD^*$
bound or resonant systems.  Such a kinematic condition is similar to
the so-called ``triangle singularity" discussed in
Refs.~\cite{Wu:2011yx,Wu:2012pg}.

The very same scenario also unavoidably leads to the appearance of a
cusp, i.e. a pronounced structure in the close vicinity of the
$S$-wave threshold. In contrast to a resonance, however, there is no
nearby pole present in the amplitude. In Ref.~\cite{Wang:2013cya} it
was argued that the location, strength and shape of the $Z_c(3900)$
signal are inconsistent with its interpretation as a cusp. Thus, an
explicit resonance is needed in addition. Still, if the $Z_c(3900)$
qualifies as a $\bar D^*D$ resonance, the mechanism described should
still lead to a significantly enhanced production rate, since it
naturally provides a large number of low-energy $\bar D^*D$ pairs.

The two-cut condition is operative in a limited kinematic range
only. As a result, the strength of the cusp as well as the number of
low energy $S$-wave $\bar DD^*$ pairs available for the formation of
the resonance, will strongly depend on the total energy of the
system. We therefore predict that, if the $Z_c(3900)$ is a resonance
produced via non-perturbative  $\bar DD^*$ interactions (a $\bar
DD^*$ resonance), its production rate should depend strongly on the
total energy of the system. In other words, even in the absence of a
pronounced cusp a hadron-hadron resonance can be produced, but in
its presence the production of a hadron-hadron resonance should be
largely enhanced. In this sense the total energy dependence of the
$Z_c(3900)$ production rate can be regarded as a diagnostic tool for
understanding its composition, if the two cut-condition is really
responsible for the copious production of $Z_c(3900)$ in the decay
of $Y(4260)$. On the contrary, if the $Z_c(3900)$ is predominantly
of tetra-quark nature, as proposed in Ref.~\cite{Faccini:2013lda},
the dependence of the production rate on the total energy of the
system should be much weaker.  This prediction can be
straightforwardly tested experimentally in $e^+e^-$ annihilations.

In this work, we identify the relative $S$-wave heavy
meson thresholds relevant for the decay of heavy vector mesons into
a pion and the isovector system of interest and discuss the possible
phenomenological implications of some of those in detail. In the end of the paper
we will also discuss briefly $P$-wave thresholds. Our analysis
should provide a path towards a better understanding of the
structure of some potential QCD exotics.

\section{Analysis of the $S$-wave singularity mechanism}

\begin{figure}[t!]\vspace{0cm}
\begin{center}
\hspace{2cm}
\includegraphics[scale=0.8]{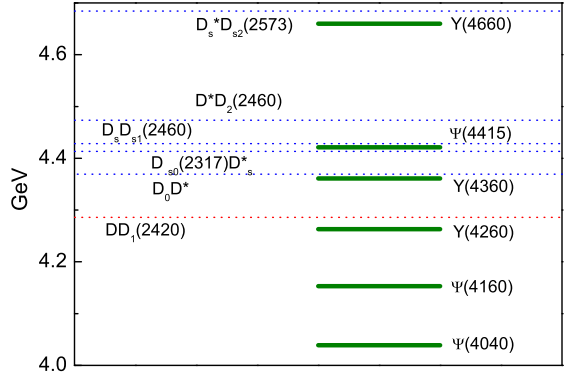}
\caption{The spectrum of vector charmonium and relative $S$-wave
open charmed thresholds. }
 \label{fig-spectrum}
\end{center}
\end{figure}

In the vector sector, the relative $S$-wave open charm thresholds
are depicted in Fig.~\ref{fig-spectrum}. Notice that the $\bar
DD_1(2420)$ system provides the first relative $S$-wave open charm
threshold. In addition, it is located near the mass position of
$Y(4260)$. It was pointed out in Ref.~\cite{Wang:2013cya}
that if the $Y(4260)$ is dominated by a
molecular $\bar DD_1$ component, one can understand the appearance of
the $Z_c(3900)$ in $e^+e^-\to Y(4260)\to J/\psi\pi\pi$ quite
naturally. In order to distinguish an explicit resonance from a cusp
effect, besides looking at the particular shape and strength of the
signal in the above mentioned reaction as done in
Ref.~\cite{Wang:2013cya}, we here explore a broader kinematic
region.

We stress that molecular states and hadron-hadron resonances cannot
be formed by broad intermediate states~\cite{Filin:2010se}.
In addition, a cusp effect will also become invisible with broad
intermediate states~\cite{complexbranchpoint}. Taking this into
account, there is only limited number of thresholds that can produce
significant cusp effects for relative $S$-wave low-momentum $\bar
DD^*$ or $\bar BB^*$ pairs, i.e.  $\bar D^{(*)} D_1(2420)$, $\bar
D^*_s D_{s0}(2317)$, $\bar D^*_s D_{s1}(2460)$, $\bar{D}^*D_2(2460)$
and $\bar{D}_s^*D_{s2}(2573)$
in the charm sector, and $\bar BB_1$ and some other corresponding
bottomed meson pairs in the bottom sector.

\begin{figure}[t!]\vspace{0cm}
\begin{center}
\hspace{2cm}
\includegraphics[scale=0.6]{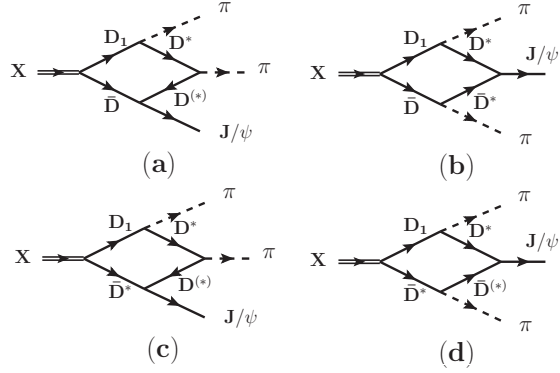}
\caption{Feynman diagrams demonstrating a vector meson $X$ with
hidden charm decays into $J/\psi\pi\pi$ via the singularity
mechanism. The Feynman diagrams in the bottomonium sector are
analogous. }
 \label{fig-Feynman-Diagram}
\end{center}
\end{figure}

\begin{figure}[t!]\vspace{0cm}
\begin{center}
\hspace{2cm}
\includegraphics[scale=0.25]{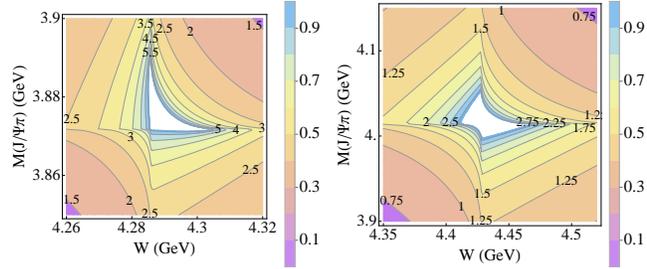}
\caption{The singularity region of $DD^*$ (left panel) and $D^*D^*$
(right panel) without considering the widths of the intermediate
mesons. The charge conjugate transitions, which behave in a similar
manner, are not included here.  The numbers in the figures are the absolute
values of three point scalar functions and the numbers in the sidebar are
their relative strengths.}
 \label{fig-DD}
\end{center}
\end{figure}

In order to demonstrate the dynamic features of the relative
$S$-wave couplings and low-momentum $\bar DD^*$ scatterings, we
employ the following Lagrangians in the calculation
\begin{eqnarray}\nonumber
\mathcal{L}_Y&=&iy(D_a^\dag Y^i \bar{D}_{1a}^{\dag i}-D_{1a}^{\dag i} Y^i \bar{D}_a^\dag)\\
&+&y\epsilon^{ijk}(D_{1a}^{\dag i}Y^k \bar{D}^{*\dag j}_a-D_a^{*\dag
j} Y^k \bar{D}_{1a}^{\dag i})+H.c.
\end{eqnarray}
for the $Y(4260)$ coupling to other $D^{(*)}$ mesons, and
\begin{eqnarray}\nonumber
\mathcal{L}_{D_1}&=&i\frac{h^\prime}{f_\pi}
[3D_{1a}^i(\partial^i\partial^j\phi_{ab})D^{*\dag
j}_b-D_{1a}^i(\partial^{j}\partial^j\phi_{ab})D_b^{*\dag
i}\\ &+&3\bar{D}_a^{*\dag
i}(\partial^i\partial^j\phi_{ab})\bar{D}_{1b}^j-\bar{D}_a^{*\dag
i}(\partial^j\partial^j\phi_{ab})\bar{D}_{1b}^i]+H.c.
\end{eqnarray}
for the $D_1$ coupling to $D^{(*)}$ and a pion. Here the $D$ ($\bar
D$) field contains the annihilation operators for the $c\bar{q}$ ($\bar{c}q$) quark configuration. The $D^*$ and $D_1$ fields are
constructed analogously. To account for the heavy quark spin
symmetry, $D$ and $D^*$ are collected into a single multiplet which
makes them share the same coupling constants  $y$ and
$h^\prime$~\cite{Casalbuoni:1996pg,Colangelo:2005gb}. The details
for the other interactions can be found in
Ref.~\cite{Cleven:2013sq}.

In the cusp kinematic region where  the intermediated states are (nearly) on shell, the exchanged charmed meson between the $J/\psi$ and a pion is far off-shell. Within such a kinematic region the propagator for the exchanged charmed meson is approximately $1/M_D^{(*)2}$ and the $\bar D^{(*)}D^*\to J/\psi \pi$ amplitude can be treated as a local function $\mathcal{F}(M(J/\psi\pi),t)$ with $M(J/\psi\pi)$ and $t$ the invariant mass of $J/\psi\pi$ and $t$-channel momentum transfer, respectively. Since $\mathcal{F}(M(J/\psi\pi),t)$ does not vary drastically within the range of $M(J/\psi\pi)$ and $t$, the four-point loop function in Fig.~\ref{fig-Feynman-Diagram} can be expressed as the following typical expression and be analyzed as a three-point function:
\begin{eqnarray}\nonumber
M&=&\int \frac{d^4 l}{(2\pi)^4}\frac{G\epsilon_{X}^i\epsilon_{J/\psi}^j(3q_1^iq_1^j-|q_1|^2\delta^{ij})\mathcal{F}(M(J/\psi\pi),t)}
{(l^0-\frac{|\vec{l}|^2}{2m_{D_1}}+i\varepsilon)(p^0-l^0-\frac{|\vec{l}|^2}{2m_{D^{(*)}}}+i\varepsilon)(l^0-q_1^0-\frac{|\vec{l}-\vec{q_1}|}{2m_{D^*}}+i\varepsilon)}\nonumber\\
&\equiv &G\epsilon_{X}^i\epsilon_{J/\psi}^j(3q_1^iq_1^j-|q_1|^2\delta^{ij})\mathcal{F}(M(J/\psi\pi),t) \mathrm{I}(m_{D_1},m_{D^{(*)}},m_{D^*},W,M(J/\psi\pi),m_\pi) \ ,
\end{eqnarray}
where $q_1$ is the three momentum of the pion connected to the initial vector charmonium through the $D_1$, $G$ is the product of all the coupling constants from different vertices and $\mathrm{I}$ is the scalar three point loop function.  
Since our focus is on the dependencies of the loops on the external parameters in order to identify the singularity regions, we set
$G=1$ and 
only use the three-point scalar function $\mathrm{I}$. This allows us to also investigate the effect of the width of the intermediate mesons.
In any physical transition,  pre-factors, which depend on the three-momentum $q_1$, 
can distort the spectra to some extend, however, the general features 
of the amplitudes persist.

\subsection{Kinematics satisfying the two-cut condition in the
vicinity of $\bar DD_1(2420)$}\label{sect-dd1}

For the final state $J/\psi\pi\pi$, the kinematics in favor of the
two-cut condition in the intermediate meson loops depends
simultaneously on both the initial mass ($W$) as well as the invariant
 mass of $J/\psi
\pi^\pm$ ($M(J/\psi\pi)$).  A $\bar D^*D$ cusp effect will be
produced by diagrams (a) and (b) of Fig.~\ref{fig-Feynman-Diagram}.
To illustrate this we show in the left panel of Fig.~\ref{fig-DD}
the modulus of results for these diagrams in the
$W$-$M(J/\psi\pi)$-plane. For simplicity in
Subsections~\ref{sect-dd1}-\ref{sect-bb}
 all intermediate mesons are treated as stable. The effect
of their widths will be discussed later in
Subsection~\ref{sect-width}.

A singularity region can be identified where the transition
amplitude is strongly enhanced and a pronounced cusp is expected
around the $\bar DD^*$ threshold region for $4.28<W<4.31$ GeV.
Unfortunately, in the preferred kinematic range there is no vector
resonance, as can be read off from Fig.~\ref{fig-spectrum}. Still,
an energy scan of the $e^+e^-$ system in this energy range would be
very valuable. Interestingly, it should be noticed that there is
still a visible enhancement even for  $W\simeq 4.26$ GeV as shown
in Fig.~\ref{fig-DD}(a), due to the strong curvature of the contour lines.
 It is this enhancement that was discussed in
Ref.~\cite{Wang:2013cya}.

In the diagrams of Fig.~\ref{fig-Feynman-Diagram}, the pion is
radiated by the narrow $D_1(2420)$ which is assigned to be the mixed
partner of the broad $D_1(2430)$ between the $^1P_1$ and $^3P_1$
states~\cite{Zhong:2008kd}. The spin symmetry demands that the
$^1P_1$ state decays into $D^*\pi$ via a $D$-wave while the $^3P_1$
decays via an $S$-wave. Thus, it is the former that is to be
identified with $D_1(2420)$,
 although some heavy quark symmetry breaking effects
are expected and may result in mixings between these two
configurations~\cite{Zhong:2008kd,Close:2005se}.

Given that the narrow $D_1(2420)$ is to be dominated by the $^1P_1$
configuration, it will introduce a different momentum and
angular dependence for
the produced pion in comparison with the so-called ``initial state
pion emission (ISPE)" proposed in
Refs.~\cite{Chen:2012yr,Chen:2011pv}. Another
distinct feature of the mechanism discussed here
compared to the ISPE is its non-local character.
 A detailed measurement of the evolution of the
$\pi\pi$ invariant mass spectra in terms of the initial $e^+e^-$
c.m. energies could shed some light on the pion emission mechanism
in the future. However, a proper theoretical treatment needs the
inclusion of the $\pi\pi$ final state interactions, which goes
beyond the scope of this paper and will be studied in a separated
work.

\begin{figure}[t!]\vspace{0cm}
\begin{center}
\hspace{-1cm}
\includegraphics[scale=0.8]{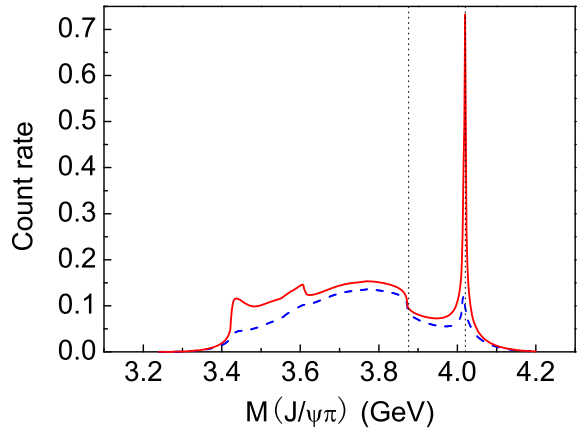}
\caption{The $J/\psi\pi$ invariant mass distribution at the center
energy $4.43~\mathrm{GeV}$ in the $J/\psi \pi\pi$ channel {\it with}
(dashed) and {\it without} (solid) the width effects for the
intermediate particles. The two vertical dotted lines denote the
$\bar DD^*$ and $\bar D^*D^*$ thresholds, respectively, from left to
right.
}
 \label{fig-JpsiPi}
\end{center}
\end{figure}

\subsection{Kinematics satisfying the two-cut condition in the vicinity of
$\bar D^*D_1(2420)$}\label{sect-dstard1}

In the energy region where the $\bar D^*D_1(2420)$ intermediate
state of diagrams (c) and (d) of Fig.~\ref{fig-Feynman-Diagram} can
be nearly on-shell, the two-cut condition is no longer satisfied for
the $\bar DD_1(2420)$ intermediate state, which means that the $\bar
DD^*$ cusp effect cannot be produced significantly in this energy
region. However, in this kinematic region the $\bar D^*D^*$ cusp can
be produced. As shown by the right panel of Fig.~\ref{fig-DD}, a
strong enhancement can be expected in the invariant mass of
$J/\psi\pi$ around the $\bar D^*D^*$ threshold for $4.42<W<4.46$
GeV. This region also extends (though somewhat less pronounced) even
down to values of $W$ as low as 4.38 GeV.

From Fig.~\ref{fig-DD} it becomes apparent that a simultaneous
appearance of cusps at both the $\bar DD^*$ and $\bar D^*D^*$
thresholds is not kinematically favored. This provides an
explanation why there was no structure near the $\bar D^*D^*$
threshold observed simultaneously with the discovery of the
$Z_c(3900)$.

In addition,  not only the strength but also the lineshape of the
cusps change rather significantly when the initial energy changes.
Such a behavior is very different from that of a resonance or a
bound state, since their pole position is independent of the initial
energy. We therefore expect that the dependence of the shape of a
near-threshold structure on the initial energy contains direct
information on the relative importance of the cusp and the resonance
pole for a particular signal.

With the (red) solid line in Fig.~\ref{fig-JpsiPi}, we show the invariant mass distribution of
$J/\psi\pi$ at $4.43$ GeV due to the processes listed in
Fig.~\ref{fig-Feynman-Diagram} --- still with all particles assumed
stable. This is the
energy region where the two-cut condition is satisfied for  $\bar D^*D^*$
and a very
pronounced  cusp occurs at this threshold. Meanwhile, since for the $\bar
DD^*$ threshold the two-cut condition is not satisfied for this
initial energy, the corresponding cusp disappears, although it is
accessible kinematically.

\subsection{Kinematics satisfying the two-cut condition in the vicinity of
$\bar B^*B_1(5721)$ and $\bar B^{(*)}B^*$}\label{sect-bb}

\begin{figure}[t!]\vspace{0cm}
\begin{center}
\hspace{2cm}
\includegraphics[scale=0.25]{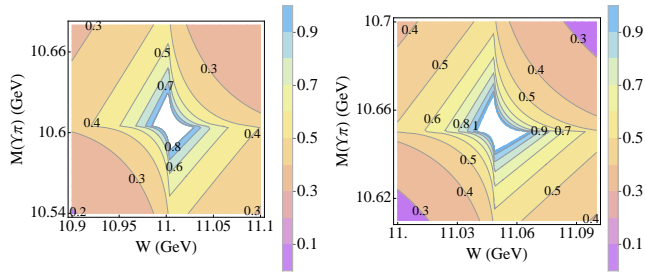}
\caption{The  singularity region of $\bar BB^*$ (left panel) and
$\bar B^*B^*$ (right panel) without considering the widths of the
intermediate mesons.  The numbers have the same meanings as those in Fig.~\ref{fig-DD}.}
 \label{fig-BB}
\end{center}
\end{figure}

The above analysis can also be applied to the bottom sector.
We present the plots showing the correlations between the initial
mass and the invariant mass of $\Upsilon(3S)\pi$ in
Fig.~\ref{fig-BB}, where the cusps caused by the $\bar BB^*$ (left
panel) or $\bar B^*B^*$ (right panel) threshold can be easily
identified. The singularity regions are very similar to those in the
charm sector except that now there is a common kinematic region that
allows those two cusps from the $\bar BB^*$ and $\bar B^*B^*$
thresholds to appear simultaneously. The main reason is that
$\Delta_{B}\equiv m_{B^*}-m_{B}=46 ~\mathrm{MeV}$ is much smaller than
$\Delta_{D}\equiv m_{D^*}-m_{D}=142 ~\mathrm{MeV}$.

\begin{figure}[t!]\vspace{0cm}
\begin{center}
\hspace{2cm}
\includegraphics[scale=0.6]{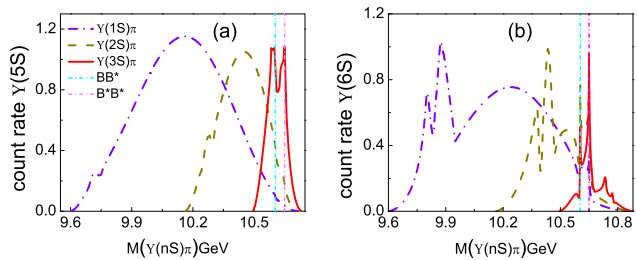}
\caption{The $\Upsilon(nS)\pi$ invariant mass distributions in (a)
$\Upsilon(5S)\to \Upsilon(nS)\pi\pi$ and (b)
$\Upsilon(6S)\to\Upsilon(nS)\pi\pi$. }
 \label{fig-UpsilonPi}
\end{center}
\end{figure}

In Fig.~\ref{fig-UpsilonPi}, we present the invariant mass
distributions of the transitions $\Upsilon(5S)\to
\Upsilon(nS)\pi\pi$ (diagram (a)) and
$\Upsilon(6S)\to\Upsilon(nS)\pi\pi$ (diagram (b)). It is interesting
to see that the production of $\Upsilon(nS)\pi\pi$ does not satisfy
the two-cut condition for the process $\Upsilon(5S)\to
\Upsilon(nS)\pi\pi$. Therefore, there are no obvious enhancements at
the $\bar BB^*$ and $\bar B^*B^*$ thresholds. In contrast, the
$\Upsilon(6S)$ lies exactly in the singularity region which makes
the two cusp peaks corresponding to the $\bar BB^*$ and $\bar
B^*B^*$ quite significant.

This result turns out to be important for understanding the nature
of $Z_b(10610)$ and $Z_b'(10650)$~\cite{Belle:2011aa}. From the
scenario studied in this work  the structures called $Z_b$ and
$Z_b'$ observed in $\Upsilon(5S)\to \Upsilon(1S, 2S)\pi\pi$ cannot
be cusps but should result from explicit resonance poles, contrary
to other claims in the literature~\cite{bugg,Chen:2012yr}.

However, for the decay of $\Upsilon(6S)$ there should be
simultaneously a large number of both $\bar BB^*$ as well as $\bar
B^*B^*$ pairs available. Therefore, if $Z_b$ and $Z_b'$ are
hadron-hadron resonances and their existences are due to the
non-perturbative $\bar B^{(*)}B^*$ interactions, their production
should be favored in the decay of $\Upsilon(6S)$. For this scenario
we therefore predict much stronger signals for these states in
$\Upsilon(6S)$ than in $\Upsilon(5S)$ decays.

\section{Influence of the width of intermediate
states}\label{sect-width}

\begin{figure}[t!]\vspace{0cm}
\begin{center}
\hspace{2cm}
\includegraphics[scale=0.25]{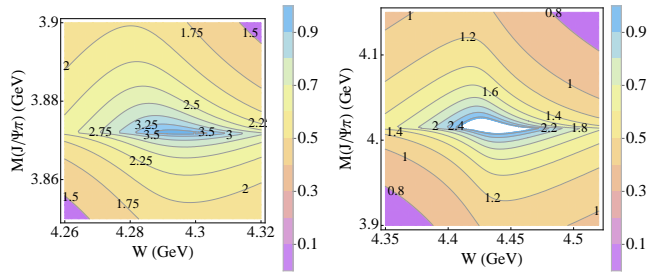}
\caption{The singularity region in the charm sector after
considering the widths of the intermediate $D_1$ and $D^*$.  Diagram
(a) is for the $\bar DD^*$ singularity region and  (b) is for $\bar
D^*D^*$. The numbers have the same meanings as those in Fig.~\ref{fig-DD}.}
 \label{fig-DD2}
\end{center}
\end{figure}

Results for the singularity regions for the loops with the widths of
the intermediate particles considered are presented in
Fig.~\ref{fig-DD2}. Taking the singularity region in the charm
sector as an example, we show the results after considering the
width of the $D_1(2420)$ with
$\Gamma_{D_1}=27~\mathrm{MeV}$~\cite{Beringer:1900zz} and
$\Gamma_{D^*}=190~\mathrm{keV}$~\cite{Seth}.
In comparison with the
results shown in Fig.~\ref{fig-DD}, we see that the cusp effects are
smeared significantly for both the $\bar DD^*$ and $\bar D^*D^*$
threshold. This becomes also clear from the dashed line in
Fig.~\ref{fig-JpsiPi}, where the $J/\psi \pi$ invariant mass
distribution is shown for $W=4.43$ GeV.

As shown by Fig.~\ref{fig-DD2}, above 4.32 GeV the structure at
$\bar DD^*$ threshold is much more like a shoulder (see also dashed
line in Fig.~\ref{fig-JpsiPi}). This should be different from the
enhancement caused by a pole structure.

\section{Analysis of the singularity mechanism in a $P$-wave transition}

\begin{figure}[t!]\vspace{0cm}
\begin{center}
\hspace{0cm}
\includegraphics[scale=0.6]{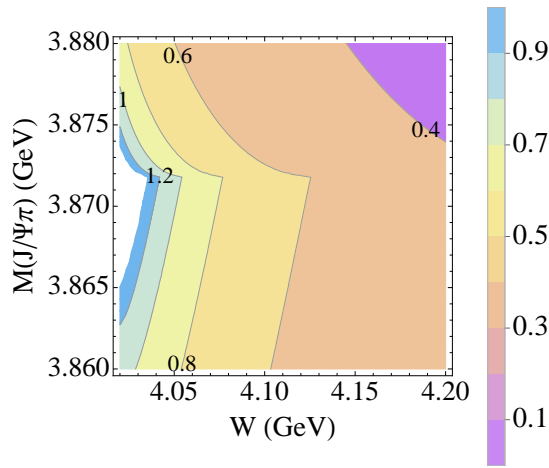}
\caption{The singularity region of heavy vector quarkonium decays
into $J/\psi\pi\pi$ via the $D^*\bar{D}^*$ intermediate loops.  The numbers have the same meanings as those in Fig.~\ref{fig-DD}.}
 \label{fig-4160}
\end{center}
\end{figure}

Recently the $Z_c(3900)$ signal is also reported in the $J/\psi\pi$
invariant mass distribution in the process $e^+e^-\to J/\psi\pi\pi$
at $4.17~\mathrm{GeV}$ by an analysis using the CLEO-c
data~\cite{Xiao:2013iha}. Although it is well below the first
$S$-wave threshold (c.f. Fig.~\ref{fig-spectrum}), there is a well
established quarkonium, $\psi(4160)$, nearby and its dominant decay
mode is $D^*\bar{D}^*$~\cite{Aubert:2009aq,Beringer:1900zz}.  Thus,
a meson loop analogous to the diagrams of
Fig.~\ref{fig-Feynman-Diagram} contains a $P$-wave vertex via the
$\psi(4160)D^*\bar D^*$ interaction. Due to the centrifugal barrier
cusps do not occur for partial waves higher than $S$-waves. Although
for higher partial waves there still is a non-analyticity, it
becomes visible in the derivative of amplitudes only~\cite{felipe}.
However, the second part of the loop still produces a cusp, as can
be seen in Fig.~\ref{fig-4160}. Although the singularity region is
now more limited in phase space, it still gives rise to some mild
enhancement at 4.17 GeV. It implies that in order to explain the
resonance signal observed by CLEO-c, an explicit resonance may be
necessary, which turns out to be consistent with our findings in the
higher energy region. Meanwhile, we anticipate that for $\psi(4040)$
even though it can give access to the $\bar DD^*$ cusp via its
strong coupling to the $\bar DD^*$
channel~\cite{Aubert:2009aq,Beringer:1900zz}, the phase space would
be extremely small and it remains to be seen if the $Z_c(3900)$ is
observable at that low energies.

\section{Summary and Perspectives}

The above analysis has identified the kinematic regions in the heavy
vector meson sectors where the relative $S$-wave heavy meson open
thresholds may play an important role when a nearby vector state
decays into a lighter quarkonium state plus two pions. It shows that
there exist mass regions that can fulfill the two-cut condition such
that the intermediate heavy meson loop can produce significant cusp
effects. The clarification of the origin of the cusps and their
evolutions with the initial masses would be important for a better
understanding of these near-threshold enhancements recently observed
in experiment, i.e. $Z_b$, $Z_b'$ and $Z_c$ etc. Based on our
analysis, we argue, that these states should not be purely due to
cusp effects if they can be observed out of the kinematics of the
singularity regions identified in this work. We further argue that
the dependence of these states on the initial energy for the
production should reveal more information on whether they can be
viewed as (predominantly) hadronic molecules or hadron-hadron
resonances, or whether they should be viewed as more complicated
structures.

The authors thank E. Eichten, F.-K. Guo, U.-G. Mei{\ss}ner, and
C.-Z. Yuan for useful discussions. This work is supported, in part,
by the National Natural Science Foundation of China (Grant Nos.
11035006 and 11121092), the Chinese Academy of Sciences
(KJCX3-SYW-N2), the Ministry of Science and Technology of China
(2009CB825200), and DFG and NSFC funds to the Sino-German CRC 110.

\end{document}